\documentclass[useAMS,usenatbib]{mn2e}
\usepackage[dvips]{graphicx}
\usepackage{longtable}
\newcommand{\chandra}{\textit{Chandra}}
\newcommand{\cxo}{\textit{Chandra X-ray Observatory}}
\newcommand{\ergcms}{\ensuremath{\mathrm{erg~cm}^{-2}~\mathrm{s}^{-1}}}
\newcommand{\Lx}{\ensuremath{L_\mathrm{x}}}
\newcommand{\Tx}{\ensuremath{T_\mathrm{x}}}
\newcommand{\Fx}{\ensuremath{F_\mathrm{x}}}
\newcommand{\kms}{\ensuremath{\mathrm{km~s}^{-1}}}
\newcommand{\err}[2]{\ensuremath{^{_{+#1}}_{^{-#2}}}}
\newcommand{\ee}[2]{\ensuremath{{#1}\!\times\!10^{#2}}}

\title[Type IIP Supernova SN 2004et]{Type IIP Supernova SN 2004et: A Multi-Wavelength Study in X-Ray, Optical and Radio}

\author[Kuntal Misra et al.]
{Kuntal Misra$^1$, Dave Pooley$^{2,\star}$, Poonam Chandra$^{3,4}$, D. Bhattacharya $^5$, \and Alak K. Ray $^6$, Ram Sagar $^1$ and Walter H. G. Lewin $^7$\\
1. Aryabhatta Research Institute of Observational Sciences, Manora Peak, Nainital, 263 129, India\\
2. Astronomy Department, University of California at Berkeley, Berkeley, CA 94720, USA\\
3. National Radio Astronomy Observatory, Charlottesville, VA 22903 \\
4. University of Virginia, Charlottesville, VA 22904\\
5. Raman Research Institute, Bangalore, 560 080, India\\
6. Tata Institute of Fundamental Research, Homi Bhabha Road, Mumbai, 400 005, India\\
7. Center for Space Research and Department of Physics, Massachusetts Institute of Technology,\\
   ~~~70 Vassar Street, Building 37, Cambridge, MA 02139-4307\\
$\star$ Chandra Fellow\\
(E-mail:kuntal{@}aries.ernet.in, dave{@}astron.berkeley.edu, pc8s{@}virginia.edu, dipankar{@}rri.res.in,
akr{@}tifr.res.in,\\
sagar{@}aries.ernet.in, lewin{@}space.mit.edu)}

\begin{document}

\date{Accepted.....; Received .....}

\pagerange{\pageref{firstpage}--\pageref{lastpage}} \pubyear{}

\maketitle

\label{firstpage}

\begin{abstract}
We present X-ray, broad band optical and low frequency radio observations of
the bright type IIP supernova SN 2004et. The \cxo\ observed the supernova at three
epochs, and the optical coverage spans a period of $\sim$ 470 days since explosion.
The X-ray emission softens with time, and we characterise the
X-ray luminosity evolution as $\Lx \propto t^{-0.4}$.  We use the observed X-ray luminosity to
estimate a mass-loss rate for the progenitor star of $\sim \ee{2}{-6}~M_\odot~\mathrm{yr}^{-1}$.
The optical light curve shows a pronounced plateau lasting for about 110 days. Temporal evolution of
photospheric radius and color temperature during the plateau phase is determined by making black body fits.
We estimate the ejected mass of $^{56}$Ni to be 0.06 $\pm$ 0.03 M$_\odot$. Using the expressions
of Litvinova \& Nad\"{e}zhin (1985) we estimate an explosion energy of
(0.98 $\pm$ 0.25) $\times 10^{51}$ erg.  We also present a single epoch radio observation of SN 2004et.
We compare this with the predictions of the model proposed by Chevalier et al. (2006).
These multi-wavelength studies suggest a main sequence progenitor mass of $\sim$ 20 M$_\odot$ for SN 2004et.
\end{abstract}

\begin{keywords}
supernovae: general - supernovae: individual: SN 2004et: techniques: photometric
\end{keywords}

\section{INTRODUCTION}
Supernovae of the type II are generally associated with regions
of star formation in spiral galaxies. They are believed to result from the
explosion triggered by core collapse of massive stars, presumably red supergiants with main sequence mass of
10$-$25 M$_{\odot}$, having a thick hydrogen envelope. Spectra of type II supernovae
show the evidence of hydrogen near maximum light. A further differentiation is proposed
for SN II based on the shape of light curve. Those which have a pronounced plateau
and remain within $\sim$ 1 mag of maximum brightness for an extended period are
termed as type IIP (plateau) while those showing a linear decline in magnitude from the peak are
termed as type IIL (linear).

The plateau duration is usually between 60$-$100 days which
is followed by an exponential tail at later epochs. The plateau phase corresponds
to a period of nearly constant luminosity due to the hydrogen recombination wave receding
through the envelope and it slowly releases the energy which was deposited by the shock
and radioactive decay.
Over the years enormous progress has been made in the study of supernovae. But the detailed
photometric and spectroscopic data especially for SN IIP are still rare.
Also the direct determination of the progenitors has been possible due to the availability of many
ground and space based archival images.
A well studied sample of SN IIP and their observational properties during the plateau phase will give
valuable information about the range
of progenitor masses, the explosion energy, the ejected mass and the amount of ejected $^{56}$Ni.

In this paper, we report the \chandra\ X-ray observations, detailed optical observations and
radio detection (at 1.4 GHz) of a type IIP supernova SN 2004et which occurred
in the spiral star burst galaxy NGC 6946 which is located at a distance of 5.5 Mpc.
Due to its proximity (z = 0.00016,) was a very promising candidate to
study the overall evolution at different wavelengths. It was the second brightest
supernova (unfiltered mag 12.8) detected in the year 2004, the brightest being a type IIP
SN 2004dj (unfiltered mag 11.2) in NGC 2403. The X-ray, optical photometric and
radio observations of SN 2004et will make it one of the best studied type IIP supernova in the
present era. Supplementing this is the progenitor identification of SN 2004et by Li et al. (2005)
using the pre-supernova images of NGC 6946 with the CFHT.

SN 2004et was discovered on September 27, 2004 by S. Moretti (Zwitter et al. 2004)
with a 0.4 m telescope.
The location of the supernova, R. A. 20$^{\rm h}$ 35$^{\rm m}$ 25$^{\rm s}$.33
and Decl. +60$^{\circ}$ 07$^{\prime}$ 17.$^{\prime\prime}$3 (J2000.0) was
247.$^{\prime\prime}$1 east and 115.$^{\prime\prime}$4 south
of the nucleus of the galaxy NGC 6946.
A high resolution spectroscopy obtained with the 1.82 m Asiago telescope on September 28
confirmed it as a type II event and the equivalent width of Na I D II lines implied
a total reddening of E(B-V) = 0.41 mag. Further observations were carried out by
Li and Filippenko (2004) with the 0.76 m Katzman Automated Imaging Telescope (KAIT).

The TAROT robotic telescope which imaged NGC 6946 frequently found no optical source at the SN location
on September 22.017.
Based on these TAROT measurements Li et al. (2005) constrain the explosion date for SN 2004et as
September 22.0 (JD 2453270.5) which we have adopted as the explosion epoch in this paper.

Li and Filippenko (2004) report that the progenitor of SN 2004et was
seen as a faint, extended source on NGC 6946 images taken with 0.9 m Kitt Peak telescope on
May 30, 1989 and
the high quality images taken with
3.6 m CFHT on August 6, 2002 show the presence of a possible progenitor within
0.$^{\prime\prime}$3 of the SN 2004et position. The estimated luminosity for these is
consistent with a massive supergiant but it is too bright and too blue for a single red supergiant.
Li et al. (2005) present a detailed analysis of the progenitor of supernova SN 2004et and
conclude that the progenitor was a yellow supergiant (15$^{+5}_{-2}$ M$_\odot$) which might have
experienced a red supergiant stage before explosion. They also mention the possibility
of the progenitor being an interacting system of a red and a blue supergiant similar to the
progenitor of supernova SN 1993J. SN 2004et progenitor is the seventh one which is directly
identified.

SN 2004et was also detected at radio wavelengths (22.4 GHz and 8.4 GHz) by Stockdale et al. (2004)
using the VLA on October 5.128 UT. No radio emission was detected on September 30.18 UT. The
radio position is in close agreement with the optical position. Beswick et al. (2004) report
the radio observations at 4.9 GHz using a subset of the MERLIN array. NGC 6946 has been the galaxy most
prolific at producing supernovae discoveries, hosting a total of 8 objects
(SN 1917A, SN 1939C, SN 1948B, SN 1968D, SN 1969P, SN 1980K, SN 2002hh and SN 2004et).
Four of these have been detected at radio wavelengths (SN 1968D, SN 1980K, SN 2002hh and SN 2004et).
SN 2004et was detected at X-ray wavelengths and was observed by $Chandra$ X-ray satellite.

We have carried out the multicolor optical photometric observations of SN 2004et from
$\sim$ 14 to 470 days after the explosion. Radio observations of SN 2004et were
carried out at a single epoch at 1.4 GHz.
Section 2 briefly discusses the X-ray, optical and radio observations.
Development of X-ray light curve and spectrum, optical photometric evolution and interpretation
of the radio observation
are discussed in the sections 3, 4 and 7 respectively. Comparison of SN 2004et with other type IIP
supernovae forms section 5, X-ray emission is discussed in section 6 whereas the conclusions form section 8
of the paper.

\section{OBSERVATIONS AND DATA REDUCTION}

\subsection{X-ray observations and \chandra\ Data Reduction}

SN~2004et was observed with \chandra\ on three occasions --- 30, 45, and 72 days since
explosion --- for $\sim$30 ks each, as part of a program to explore the X-ray properties of
core-collapse SNe (PI: Lewin); details of the observations can be found in Table~\ref{tab:xray}.
An initial report of the observations was made by Rho et al. (2007).  All data were taken with the
Advanced CCD Imaging Spectrometer (ACIS) with an integration time of 3.2 s per frame.  The telescope
aimpoint was on the back-side illuminated S3 chip, and the data were telemetered to the ground in ``faint'' mode.

Data reduction was performed using the CIAO 3.3 software provided by the Chandra X-ray
Center\footnote{{http://asc.harvard.edu}}.  The data were reprocessed using the CALDB 3.2.2 set of
calibration files (gain maps, quantum efficiency, quantum efficiency uniformity, effective area)
including a new bad pixel list made with the acis\_run\_hotpix tool.  The reprocessing was done
without including the pixel randomization that is added during standard processing.  This omission
slightly improves the point spread function.  The data were filtered using the standard ASCA grades
(0, 2, 3, 4, and 6) and excluding both bad pixels and software-flagged cosmic ray events.
This is the standard grade filtering for ACIS imaging observations recommended by the \chandra X-ray
Center; this filtering optimises the signal-to-background ratio.
Intervals of strong background flaring were searched for, but none were found.

\subsection{Optical Observations and Data Analysis}
We have carried out the broad band UBVR$_{c}$I$_{c}$ optical photometric observations of SN 2004et
at 29 epochs during October 06, 2004 to January 01, 2005.
Further observations were precluded due to its proximity to the Sun.
The observations were continued again in October 2005. We could
observe SN 2004et at 24 epochs during October 13, 2005 to January 05, 2006. We present, here, an
extended coverage of SN 2004et at 53 epochs spanning over 470 days since the explosion on
September 22, 2004.
All the observations were carried out using a 2048 $\times$ 2048 CCD camera mounted at the f/13
Cassegrain focus of the 1.04-m Sampurnanand Telescope (ST) at
Aryabhatta Research Institute of Observational Sciences (ARIES), Nainital.
One pixel of the
CCD chip corresponds to a square of 0\arcsec.38 side, and the entire chip covers a field of
13\arcmin $\times$ 13\arcmin on the sky. The gain and read out noise of the CCD camera are 10 electrons
per analogue-to-digital unit (ADU) and 5.3 electrons respectively.

We observed the Landolt (1992) standard PG 0231+051 field and the supernova in UBVR$_{c}$I$_{c}$ filters
on November 14, 2004 for photometric calibration during good photometric sky conditions. Several bias and
twilight flat frames were observed with the CCD camera to calibrate the images using standard techniques.
The bias subtracted, flat fielded and cosmic ray removed images were used for further analysis.

Image processing was done using softwares IRAF, MIDAS and DAOPHOT-II. The values of the atmospheric
extinction on the night of 14/15 November 2004 determined from the observations of PG 0231+051
bright stars are 0.53, 0.24, 0.14, 0.10 and 0.07 mag in U, B, V, R$_{c}$ and I$_{c}$ filters
respectively. The 6 standard stars in the PG 0231+051 field cover a range of -0.329 $< (B-V) <$ 1.448
in color and of 12.770 $< V <$ 16.110 in brightness. This gives the following transformation coefficients\\

\noindent
$u_{ccd}$ = $U - (0.061 \pm 0.005) (U-B) + (6.937 \pm 0.004)$\\
$b_{ccd}$ = $B - (0.015 \pm 0.004) (B-V) + (4.674 \pm 0.004)$\\
$v_{ccd}$ = $V - (0.011 \pm 0.006) (B-V) + (4.223 \pm 0.006)$\\
$r_{ccd}$ = $R - (0.001 \pm 0.011) (V-R) + (4.136 \pm 0.007)$\\
$i_{ccd}$ = $I - (0.023 \pm 0.018) (R-I) + (4.619 \pm 0.012)$\\

\noindent
where $U, B, V, R, I$ are standard magnitudes and $u_{ccd}, b_{ccd}, v_{ccd}, r_{ccd}, i_{ccd}$
represent the instrumental magnitudes normalized for 1 s of exposure and corrected for atmospheric
extinction.
The color coefficients, zero-points and errors in them were determined by fitting least
square linear regressions to the data points.
Using the transformation coefficients obtained, $UBVRI$ photometric magnitudes
of 11 secondary stars were determined in the SN 2004et field and their average values are listed in
Table \ref{calib_stars}. The uncertainties are indicated in the parentheses.

\begin{table*}
\caption{The identification number, standard $V, (U-B), (B-V), (V-R)$ and $(R-I)$ photometric magnitudes
of the secondary stars in the SN 2004et field are given. Uncertainties are indicated in the parenthesis.}
\smallskip
\begin{center}
\begin{tabular}{@{}cccccc}\hline \hline
ID& $V$ & $U-B$ & $B-V$ & $V-R$ & $R-I$\\
&(mag)&(mag)&(mag)&(mag)&(mag)\\ \hline
&&&&&\\
1 &17.10 (0.012)&0.97 (0.114)&1.21 (0.021)&0.67 (0.022)&0.74 (0.032)\\
2 &18.48 (0.015)&0.02 (0.108)&0.92 (0.031)&0.52 (0.020)&0.65 (0.055)\\
3 &17.66 (0.013)&0.02 (0.038)&0.77 (0.022)&0.38 (0.045)&0.61 (0.059)\\
4 &17.39 (0.011)&0.31 (0.035)&0.94 (0.019)&0.46 (0.013)&0.68 (0.029)\\
5 &17.46 (0.013)&0.20 (0.045)&0.84 (0.022)&0.42 (0.014)&0.65 (0.030)\\
6 &14.15 (0.009)&1.23 (0.013)&1.36 (0.012)&0.74 (0.010)&0.78 (0.023)\\
7 &14.28 (0.009)&1.06 (0.012)&1.28 (0.012)&0.65 (0.010)&0.82 (0.023)\\
8 &14.78 (0.009)&0.13 (0.010)&0.74 (0.013)&0.44 (0.010)&0.53 (0.023)\\
9 &13.80 (0.009)&0.09 (0.010)&0.63 (0.012)&0.39 (0.010)&0.40 (0.023)\\
10&14.71 (0.009)&0.33 (0.010)&0.82 (0.012)&0.49 (0.010)&0.46 (0.023)\\
11&14.73 (0.010)&0.62 (0.012)&1.07 (0.013)&0.59 (0.011)&0.67 (0.023)\\
&&&&&\\
\hline
\end{tabular}
\end{center}
\label{calib_stars}
\end{table*}

Li et al. (2005) present the magnitudes of 8 secondary standards in the field of SN 2004et. The zero-point
differences, based on comparison of 7 secondary standards in the field of SN 2004et, between our photometry
and that of Li et al. (2005) are
0.033 
, -0.001 
and 0.016 
in $B, V$ and $R$ filters respectively. This shows that our photometric calibration is secure.
We cannot compare the $U$ and $I$ band magnitudes as
Li et al. (2005) mention the $U$ and $I$ band magnitudes for only 2 secondary stars.
differences are based on the comparison of the 7 secondary stars in the field of SN 2004et.

Several short exposures, with exposure time varying from 60 sec to 1800 sec, were taken to image
SN 2004et. The data have been binned in 2 $\times$ 2 pixel$^2$ because the data are oversampled
and data analysis is simplified having a higher signal-to-noise ratio per pixel with fewer pixels.
The supernova was differentially calibrated with respect to the secondary stars 6, 7, 8 and 9 listed
in Table \ref{calib_stars}.
A full compilation of SN 2004et magnitudes in $UBVRI$ filters derived in this way along with errors
is presented in Table \ref{sn2004et_mag}.

\begin{figure}
\includegraphics[width=84mm]{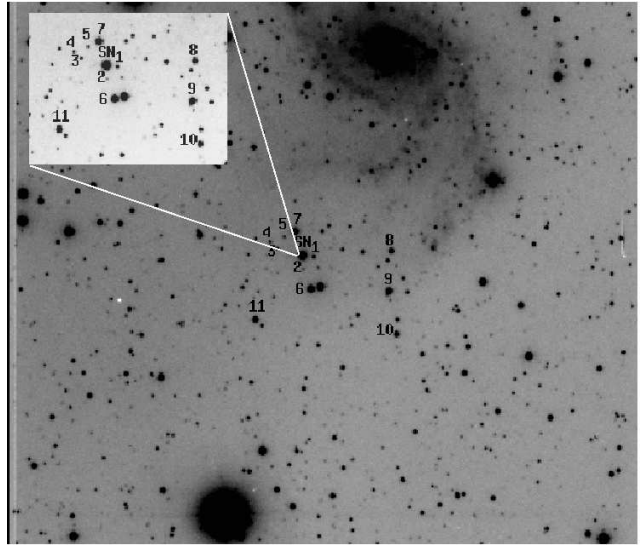}
\caption{SN 2004et field observed with the 1.04-m Sampurnanand Telescope at ARIES, Nainital.
Inset shows the location of supernova and marked are the comparison stars.}
\label{sn2004et_chart}
\end{figure}

\begin{table*}
\caption{UBV${\rm R_{c}}$ and ${\rm I_{c}}$ magnitudes of SN 2004et along with
errors, Julian date and mid UT of observations are listed.}
\smallskip
\begin{center}
\begin{tabular}{@{}ccccccc}\hline \hline
Date& Time &$U$ &$B$ &$V$ &$R_{c}$ & $I_{c}$ \\
(UT)&(JD)&(mag)&(mag)&(mag)&(mag)&(mag)\\
\hline
&&& & & & \\
2004 10 06.630&2453285.1296&12.44 $\pm$ 0.016&12.95 $\pm$ 0.014&12.60 $\pm$ 0.019&12.27 $\pm$ 0.013&11.94 $\pm$ 0.019\\
2004 10 13.616&2453292.1162&                 &12.88 $\pm$ 0.032&12.50 $\pm$ 0.027&12.16 $\pm$ 0.027&11.88 $\pm$ 0.029\\
2004 10 15.584&2453294.0844&                 &13.20 $\pm$ 0.010&12.62 $\pm$ 0.014&12.22 $\pm$ 0.016&11.91 $\pm$ 0.015\\
2004 10 16.603&2453295.1031&13.15 $\pm$ 0.011&13.21 $\pm$ 0.013&12.62 $\pm$ 0.013&12.23 $\pm$ 0.020&11.88 $\pm$ 0.015\\
2004 10 17.552&2453296.0517&13.26 $\pm$ 0.018&13.26 $\pm$ 0.010&12.63 $\pm$ 0.026&12.24 $\pm$ 0.021&11.87 $\pm$ 0.031\\
2004 10 18.573&2453297.0726&13.40 $\pm$ 0.013&13.31 $\pm$ 0.013&12.66 $\pm$ 0.021&12.23 $\pm$ 0.030&11.88 $\pm$ 0.024\\
2004 10 19.607&2453298.1072&13.50 $\pm$ 0.023&13.37 $\pm$ 0.016&12.63 $\pm$ 0.019&12.21 $\pm$ 0.036&11.89 $\pm$ 0.031\\
2004 10 20.568&2453299.0677&13.63 $\pm$ 0.010&13.41 $\pm$ 0.013&                 &12.23 $\pm$ 0.019&11.90 $\pm$ 0.018\\
2004 10 21.554&2453300.0543&13.72 $\pm$ 0.012&13.47 $\pm$ 0.009&12.67 $\pm$ 0.014&12.22 $\pm$ 0.018&11.88 $\pm$ 0.020\\
2004 10 22.560&2453301.0596&13.81 $\pm$ 0.012&13.52 $\pm$ 0.006&12.69 $\pm$ 0.016&12.25 $\pm$ 0.014&11.89 $\pm$ 0.017\\
2004 11 03.570&2453313.0698&14.77 $\pm$ 0.007&13.96 $\pm$ 0.003&12.84 $\pm$ 0.004&12.33 $\pm$ 0.006&11.91 $\pm$ 0.008\\
2004 11 04.542&2453314.0422&14.83 $\pm$ 0.007&13.99 $\pm$ 0.008&12.85 $\pm$ 0.008&12.32 $\pm$ 0.009&11.92 $\pm$ 0.008\\
2004 11 05.570&2453315.0699&14.89 $\pm$ 0.006&14.02 $\pm$ 0.006&12.87 $\pm$ 0.007&12.33 $\pm$ 0.010&11.90 $\pm$ 0.014\\
2004 11 06.548&2453316.0476&14.95 $\pm$ 0.010&14.04 $\pm$ 0.007&12.87 $\pm$ 0.010&12.34 $\pm$ 0.010&11.90 $\pm$ 0.013\\
2004 11 08.550&2453318.0504&15.06 $\pm$ 0.005&14.08 $\pm$ 0.004&12.89 $\pm$ 0.006&12.35 $\pm$ 0.009&11.90 $\pm$ 0.008\\
2004 11 13.550&2453323.0499&15.32 $\pm$ 0.038&14.23 $\pm$ 0.011&12.93 $\pm$ 0.005&12.35 $\pm$ 0.014&11.90 $\pm$ 0.012\\
2004 11 14.543&2453324.0431&15.37 $\pm$ 0.010&14.21 $\pm$ 0.003&12.93 $\pm$ 0.010&12.35 $\pm$ 0.008&11.90 $\pm$ 0.009\\
2004 11 16.538&2453326.0377&15.48 $\pm$ 0.011&14.24 $\pm$ 0.003&12.94 $\pm$ 0.004&12.36 $\pm$ 0.008&11.89 $\pm$ 0.012\\
2004 11 18.541&2453328.0415&15.62 $\pm$ 0.012&14.31 $\pm$ 0.007&12.97 $\pm$ 0.013&12.37 $\pm$ 0.012&11.92 $\pm$ 0.010\\
2004 12 02.572&2453342.0717&16.18 $\pm$ 0.018&14.52 $\pm$ 0.004&13.04 $\pm$ 0.003&12.40 $\pm$ 0.004&11.91 $\pm$ 0.010\\
2004 12 03.587&2453343.0867&16.16 $\pm$ 0.015&14.54 $\pm$ 0.004&13.05 $\pm$ 0.003&12.41 $\pm$ 0.005&11.92 $\pm$ 0.005\\
2004 12 04.574&2453344.0742&16.22 $\pm$ 0.013&14.56 $\pm$ 0.011&13.06 $\pm$ 0.016&                 &11.92 $\pm$ 0.030\\
2004 12 12.539&2453352.0389&16.54 $\pm$ 0.017&14.69 $\pm$ 0.004&13.11 $\pm$ 0.007&12.44 $\pm$ 0.007&11.92 $\pm$ 0.010\\
2004 12 13.535&2453353.0353&                 &14.71 $\pm$ 0.005&13.14 $\pm$ 0.003&12.48 $\pm$ 0.004&                 \\
2004 12 15.578&2453355.0783&                 &14.73 $\pm$ 0.002&13.15 $\pm$ 0.001&12.47 $\pm$ 0.002&11.96 $\pm$ 0.002\\
2004 12 16.551&2453356.0512&16.63 $\pm$ 0.031&14.76 $\pm$ 0.003&13.15 $\pm$ 0.002&12.48 $\pm$ 0.004&11.93 $\pm$ 0.008\\
2004 12 18.560&2453358.0597&                 &14.80 $\pm$ 0.003&13.18 $\pm$ 0.002&12.49 $\pm$ 0.003&11.98 $\pm$ 0.003\\
2004 12 22.546&2453362.0456&                 &                 &                 &12.50 $\pm$ 0.012&11.99 $\pm$ 0.009\\
2005 01 02.560&2453371.0601&                 &                 &                 &                 &12.10 $\pm$ 0.006\\
2005 10 13.615&2453657.1147&                 &19.15 $\pm$ 0.047&18.32 $\pm$ 0.020&17.31 $\pm$ 0.018&16.84 $\pm$ 0.018\\
2005 10 14.615&2453658.1147&                 &19.15 $\pm$ 0.047&18.31 $\pm$ 0.020&17.31 $\pm$ 0.018&16.84 $\pm$ 0.017\\
2005 10 27.560&2453671.0601&                 &19.43 $\pm$ 0.038&18.41 $\pm$ 0.018&17.50 $\pm$ 0.014&                 \\
2005 10 29.576&2453673.0760&                 &19.45 $\pm$ 0.028&18.51 $\pm$ 0.017&17.59 $\pm$ 0.017&                 \\
2005 11 02.598&2453677.0980&                 &19.48 $\pm$ 0.019&18.54 $\pm$ 0.020&17.59 $\pm$ 0.015&                 \\
2005 11 04.544&2453679.0438&                 &19.53 $\pm$ 0.031&18.55 $\pm$ 0.014&17.61 $\pm$ 0.024&17.15 $\pm$ 0.019\\
2005 11 05.557&2453680.0571&20.26 $\pm$ 0.069&19.49 $\pm$ 0.022&18.59 $\pm$ 0.015&17.65 $\pm$ 0.015&17.16 $\pm$ 0.020\\
2005 11 06.577&2453681.0770&20.27 $\pm$ 0.081&19.53 $\pm$ 0.018&18.61 $\pm$ 0.017&17.70 $\pm$ 0.014&17.23 $\pm$ 0.017\\
2005 11 07.558&2453682.0576&                 &19.54 $\pm$ 0.026&18.61 $\pm$ 0.017&17.67 $\pm$ 0.016&17.21 $\pm$ 0.016\\
2005 11 09.530&2453684.0298&                 &                 &18.70 $\pm$ 0.026&17.74 $\pm$ 0.019&17.20 $\pm$ 0.037\\
2005 11 22.561&2453697.0610&                 &19.60 $\pm$ 0.037&18.78 $\pm$ 0.021&17.91 $\pm$ 0.018&                 \\
2005 11 23.560&2453698.0598&                 &19.62 $\pm$ 0.039&18.79 $\pm$ 0.024&17.92 $\pm$ 0.015&                 \\
2005 11 24.551&2453699.0512&                 &19.66 $\pm$ 0.033&18.83 $\pm$ 0.020&17.96 $\pm$ 0.019&                 \\
2005 12 03.600&2453708.1001&                 &                 &18.94 $\pm$ 0.026&18.07 $\pm$ 0.020&17.54 $\pm$ 0.031\\
2005 12 05.570&2453710.0699&20.41 $\pm$ 0.067&19.80 $\pm$ 0.030&18.99 $\pm$ 0.013&18.12 $\pm$ 0.014&17.66 $\pm$ 0.015\\
2005 12 06.545&2453711.0447&                 &19.79 $\pm$ 0.023&18.90 $\pm$ 0.025&18.16 $\pm$ 0.022&17.64 $\pm$ 0.036\\
2005 12 08.577&2453713.0766&20.66 $\pm$ 0.244&19.83 $\pm$ 0.040&19.04 $\pm$ 0.022&18.15 $\pm$ 0.021&17.64 $\pm$ 0.026\\
2005 12 09.556&2453714.0565&                 &19.85 $\pm$ 0.044&19.03 $\pm$ 0.029&18.11 $\pm$ 0.020&17.68 $\pm$ 0.026\\
2005 12 18.553&2453723.0534&                 &                 &19.11 $\pm$ 0.023&18.29 $\pm$ 0.014&                 \\
2005 12 19.554&2453724.0538&                 &                 &19.11 $\pm$ 0.038&                 &                 \\
2005 12 20.567&2453725.0674&                 &                 &19.01 $\pm$ 0.024&18.23 $\pm$ 0.022&17.73 $\pm$ 0.047\\
2005 12 21.550&2453726.0502&                 &                 &19.14 $\pm$ 0.042&18.33 $\pm$ 0.019&17.86 $\pm$ 0.051\\
2006 01 04.552&2453738.0524&                 &                 &19.38 $\pm$ 0.037&18.59 $\pm$ 0.024&                 \\
2006 01 05.556&2453739.0561&                 &20.22 $\pm$ 0.099&19.35 $\pm$ 0.037&18.58 $\pm$ 0.036&                 \\
&&& & & &\\
\hline
\end{tabular}
\end{center}
\label{sn2004et_mag}
\end{table*}

\subsection{GMRT Radio Observations and Data Analysis}
We observed SN 2004et with the Giant Meterwave Radio Telescope (GMRT)
on January 02, 2005. GMRT is a synthesis array radio
telescope consisting of 30 dish antennae of diameter 45 m each. The
observations in a frequency band centered at 1390 MHz, with a bandwidth
of 16 MHz (divided into a total 128 frequency channels).
The total no of antennae available for our observations were 27.
The total time spent on the
observation was 4 hours, out of which we spent 2.5 hours tracking the
target source Field of View (FoV). 3C48 was used as a flux calibrator.
The flux density of 3C48 was derived using Baars formulation
(Baars et al. 1977) extrapolated to the observed frequency. For phase calibration,
we used the source from VLA catalogue 2022+616. Sources 3C48 and 2022+616
were also used for bandpass calibration. 3C48
was observed twice, for 10 minutes each once at the start of the observations and once towards the
end.
The phase calibrator were observed for 6 minutes after every 25 minutes
observation of the supernova FoV. Observation of the phase calibrator
at small time intervals was important not only for tracking the
instrumental phase and gain drifts, atmospheric and ionospheric gain
and phase variations, but also for monitoring the
quality and sensitivity of the data and for spotting occasional gain
and phase jumps.

The data was analysed using Astronomical Image Processing System (AIPS) software developed
by National Radio Astronomy Observatory (NRAO). One clean channel was
picked up and VPLOT was used to look for dead antennae and obviously bad baselines.
Bad antennae and corrupted data were removed using standard AIPS routines.
Data were then calibrated and images of the fields
were formed by Fourier inversion and CLEANing using AIPS task ``IMAGR''. For 1390 MHz
observations, the bandwidth smearing effect was negligible and imaging was done after
averaging 100 central frequency channels. The whole field was divided into 2 subfields
to take care of the wide field imaging. We performed a few rounds of phase self calibration
to remove the phase variations due to the bad weather and related causes and improved
the images a good deal.

\begin{figure}
\includegraphics[width=84mm]{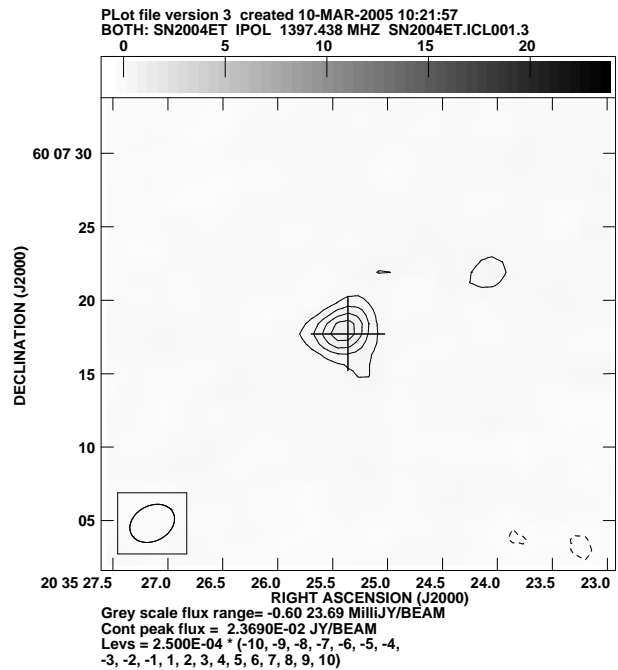}
\caption{Radio contour map of SN 2004et obtained with the GMRT on Jan 02, 2005.
The cross shows the optical position of the SN 2004et.}
\label{SN_GMRT}
\end{figure}

The flux density of 3C48 using Baars formulation is 16.15 Jy.
The flux density of the phase calibrator was $1.82 \pm 0.07$ Jy.
AIPS task
``JMFIT'' was used to fit the Gaussian at the SN position and obtain the
flux density. The flux density of SN 2004et was found to be
$ 1.24 \pm 0.24$ mJy. The map rms of the supernova FoV was
90 $\mu$Jy and the map resolution was $ 3.20'' \times 2.40''$.
The J2000 position of the SN 2004et obtained from our GMRT observations in
RA and Dec was $20^h 35^m 25^s40 \pm 0.02$ and $ 60^o 07' 17''90 \pm  0.12$ respectively,
consistent with the optical position of the supernova.
Figure \ref{SN_GMRT} shows the radio contour map of SN 2004et FoV.

\section{X-Ray Analysis}
\subsection{X-ray spectra}

We extracted counts and spectra in the 0.5--8 keV bandpass from the location of the SN using
ACIS Extract version 3.107 (Broos et al. 2002).  The source extraction region was chosen to enclose
 90\% of the point spread function at 1.5 keV, which resulted in a roughly circular region of
 $\sim$1\arcsec\ radius.  We obtained 204, 153, and 156 counts for the three observations.
Our background region was a source-free annulus centered on the SN with an inner radius of
6\arcsec\ and an outer radius of 27\arcsec.  This background region contained $\sim$100 counts
in each observation; we therefore expect the source extraction region to contain 0.14 background
counts on average.  Response files were made using ACIS Extract, which calls the CIAO tools mkacisrmf and mkarf.

We grouped the spectra to contain at least 15 counts/bin and fit them in Sherpa (Freeman et al. 2001)
using standard $\chi^2$ statistics.  We fit two simple models: an absorbed power law and an absorbed mekal
 plasma with elemental abundances allowed to vary.  In the mekal fits, if an elemental abundance was
consistent with solar, i.e., its 1\,$\sigma$ confidence interval included the solar value, we froze that
abundance and refit in order to find any potential non-solar abundances.  In all cases (power law and mekal),
the absorbing column was consistent with the value of $n_H = 2.3 \times 10^{21}~\mathrm{cm}^{-2}$ inferred
from the measurement of $E(B-V)=0.41\,\mathrm{mag}$ (Zwitter et al. 2004), using the conversion of
$5.55 \times 10^{21}~\mathrm{cm}^{-2}~\mathrm{mag}^{-1}$ (Predehl \& Schmitt 1995).
We therefore froze the column
density at this value and refit the spectra.  The results can be found in Table~\ref{tab:xray}.  We also give
the unabsorbed flux over the \chandra\ bandpass; the uncertainty in this flux estimate is derived from the
uncertainty in the model normalization, found via the projection command in Sherpa.

\begin{table*}
\setlength{\tabcolsep}{3pt}
\caption{\chandra\ observation details and spectral fit parameters for either a power-law model or mekal plasma
model.}
\smallskip
\begin{center}
\begin{tabular}{llllllllllll}\hline\hline
ObsID& Date (UT)     & Exp     & T        & PL Index          & \Fx\ [0.5--8\,keV]            & $\chi^2$/dof &&
Mekal kT          & \Fx\ [0.5--8\,keV]                & $\chi^2$/dof & Non-solar Abund.\\
     &               & (s)     & (d)      &                   & (\ergcms)                     &              &&
(keV)             & (\ergcms)                         &              &                 \\ \hline
4631 & 2004-Oct-22.2 & 29740.2 & 30.2     & 1.3\err{0.1}{0.1} & $8.3\err{0.9}{0.9}\times10^{-14}$  & 11.5/12 &&
20\err{11}{6}     & $8.2\err{0.6}{0.6}\times10^{-14}$ & 11.9/12      & --- \\
4632 & 2004-Nov-06.0 & 27980.4 & 45.0     & 1.6\err{0.2}{0.2} & $5.8\err{0.7}{0.7}\times10^{-14}$  & 17.2/8  &&
3.9\err{1.0}{0.8} & $6.4\err{1.0}{1.0}\times10^{-14}$ & 4.5/4        & O, Si, Ca, Fe\\
4633 & 2004-Dec-03.5 & 26615.5 & 72.5     & 1.9\err{0.2}{0.2} & $5.9\err{0.6}{0.6}\times10^{-14}$  & 5.9/9   &&
2.8\err{0.7}{0.4} & $5.5\err{0.7}{0.7}\times10^{-14}$ & 2.2/8        & Ca \\ \hline
\end{tabular}
\end{center}
\label{tab:xray}
Notes: T = time since explosion, taken to be 2004-Sep-22.0 (Li et al. 2005); PL = power law; all flux
es are unabsorbed; confidence intervals are 1\,$\sigma$
\end{table*}

\begin{figure*}
\includegraphics[width=\textwidth]{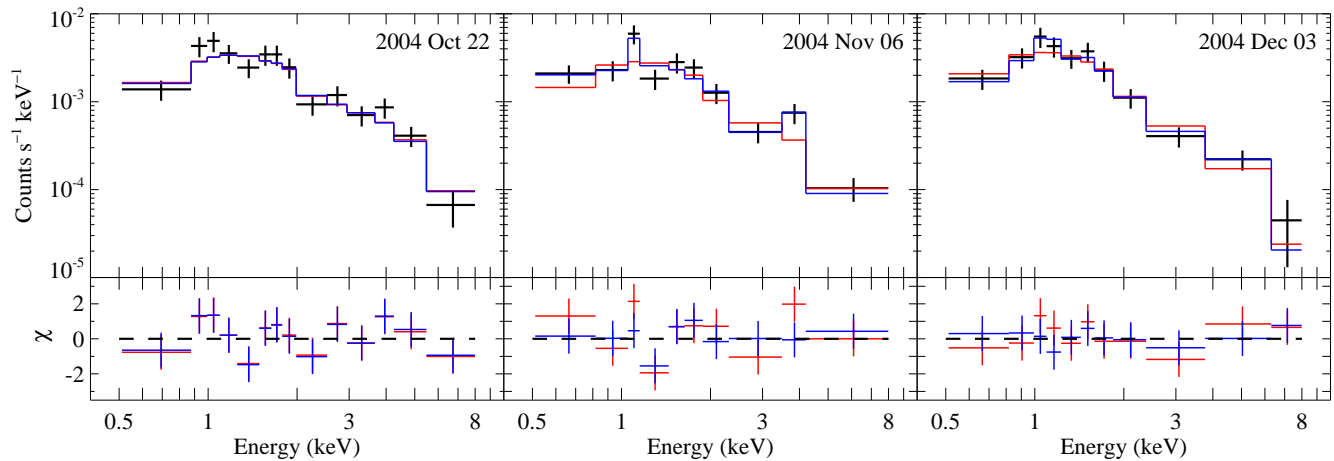}
\caption{\chandra\ spectra of SN 2004et with power law fits (red) and mekal plasma fits (blue). The residuals are plotted in the bottom panels as $(\rm{data}-\rm{model})/\rm{error}$ with vertical bars of length $\pm$1}
\label{fig:xrayspec}
\end{figure*}

Although these models may be too simplistic, they both show that the X-ray spectrum softens with time,
and they are both in fairly good agreement on the unabsorbed flux.  Given the low statistical quality
of the spectra, more complex models are not warranted.  We note that, although the data do not have much
 power to distinguish different spectral models, both the power law model and the Mekal model with solar
abundances provide statistically unacceptable fits to the second observation (without letting the abundances
vary, the Mekal fit had a $\chi^2$/dof = 17.4/8). The discrepancy between the data and these models could be
due to emission from O, Fe, Si, and Ca near 0.7, 1.1, 2.1, and 3.8 keV, respectively, as indicated by the
 mekal spectra fits with these abundances allowed to vary (see Figure~\ref{fig:xrayspec}).  However, this
is not conclusive.  The best-fit abundances are
$\rm{O}=56\err{35}{23}$, $\rm{Si}=25\err{20}{16}$, $\rm{Ca}=275\err{112}{109}$, and $\rm{Fe}=11\err{7}{4}$,
 all values with respect to solar.

Rho et al. 2007 reported that the first observation is best fitted by a ``thermal model'' with a
temperature of $kT=1.3$ keV and a lower limit of 0.5 keV.  They also report a significantly enhanced
column of $n_H=1\times10^{22}~\mathrm{cm}^{-2}$.  However, our Mekal model prefers a much higher temperature
for the first observation, $kT=20$ keV, with a lower limit of 14 keV, and the index of our power law model
 also suggests a harder spectrum than a 1.3 keV plasma would be expected to produce.  We are unable to reproduce
the low temperature found by Rho et al. 2007, with either the Mekal model or Raymond-Smith or thermal bremsstrahlung models.
 An absorbed Raymond-Smith model (with the column density allowed to vary) gives a temperature of
$kT=22\err{42}{13}$ keV (with $n_H=2.3\err{0.6}{0.5}\times10^{21}~\mathrm{cm}^{-2}$).
An absorbed bremsstrahlung model gives a temperature of $kT=14\err{\infty}{4}$ keV
(with $n_H=2.5\err{0.6}{0.5}\times10^{21}~\mathrm{cm}^{-2}$).  The reason for this discrepancy
 between our results and those of Rho et al. 2007\ are unclear, but their thermal model may have a fairly
 different distribution of flux than the ones tried here..

\subsection{X-ray Evolution}

\begin{figure}
\includegraphics[width=\columnwidth]{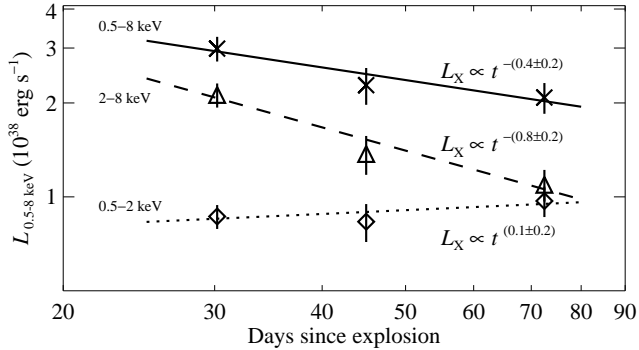}
\caption{\chandra\ light curves of SN 2004et with power law fits.  The solid line is the best fit power law for the full 0.5--8 keV luminosities (represented by $\times$).  The dashed line is for the hard band (2--8 keV) luminosities (represented by triangles).  The dotted line is for the soft band (0.5--2 keV) luminosities (represented by diamonds).}
\label{fig:xraylc}
\end{figure}

We characterize the X-ray luminosity evolution as a power law in time, $\Lx \propto t^{-\alpha}$.
We use the average of the fluxes from both models as the best estimate of the flux from each observation.
 Using a distance of 5.5 Mpc, the X-ray luminosities are \ee{(3.0\pm0.3)}{38}, \ee{(2.3\pm0.3)}{38}, and
\ee{(2.1\pm0.2)}{38} erg~$\rm{s}^{-1}$.  The best fit $\alpha$ is $0.4\pm0.2$.  The X-ray light curve is
 plotted in Figure~\ref{fig:xraylc}.  The spectral evolution can be seen by separately plotting the soft band
(0.5--2 keV) and hard band (2--8 keV) fluxes (Figure~\ref{fig:xraylc}).  The soft X-ray luminosity is nearly constant,
 while the hard band decays with a power law index of $\alpha_\mathrm{hard}=0.8\pm0.2$.  This is very different than
the evolution of the type IIP 1999em observed by \chandra\ (Pooley et al. 2002); in that case, both the hard and soft bands
 were seen to decay.

\begin{figure}
\includegraphics[width=\columnwidth]{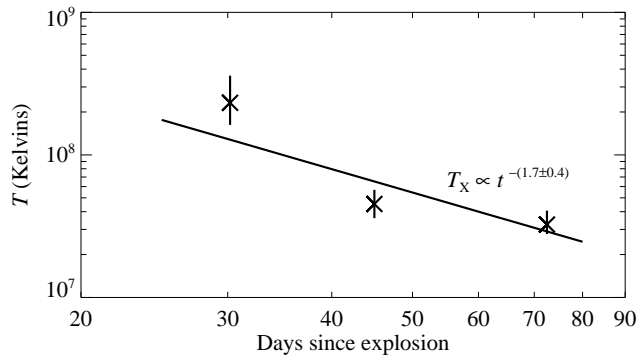}
\caption{X-ray temperature evolution of SN 2004et with power law fit.}
\label{fig:xraytemp}
\end{figure}

We further explore the spectral evolution by examining the temperature evolution, as estimated by the
single-temperature Mekal models (Table~\ref{tab:xray}).  We plot the X-ray temperature as a function of time
in Figure~\ref{fig:xraytemp}.  The evolution can be roughly characterized as a power law in time,
$\Tx \propto t^{-\beta}$, with a best fit $\beta=1.7\pm0.4$.

\section{Photometric Evolution}
In this section we present the results of our photometric observations. Since this supernova is
located 247.$^{\prime\prime}$1 east and 115.$^{\prime\prime}$4 south of the nucleus of
NGC 6946, the contribution from the galaxy is negligible in determining the supernova magnitudes.

\subsection{Reddening towards SN 2004et}
The high resolution spectra obtained by Zwitter et al. (2004) shows a relatively featureless spectrum
with very broad, low contrast H$\alpha$ emission but the presence of sharp interstellar Na I
absorption lines. The equivalent width of Na I DII lines corresponds to an estimated total
reddening of E(B-V) = 0.41 mag. The galactic extinction towards SN 2004et is E(B-V) = 0.34 mag
using the reddening maps of Schlegel et al. (1998) which corresponds to A$_{V}^{gal}$ = 1.06 mag for
a R$_{V}$ = 3.1. But the total extinction is E(B-V) = 0.41 mag, thus the rest of it may be
attributed due to the host galaxy. Li et al. (2005) adopt a total reddening of
E(B-V) = 0.41 $\pm$ 0.07 mag towards SN 2004et. The lower limit of this corresponds to no
reddening due to the host galaxy. We adopt a total reddening (galactic + host) of
E(B-V) = 0.41 $\pm$ 0.07 mag for our further analysis.

\subsection{Light Curves}
\begin{figure}
\includegraphics[width=94mm]{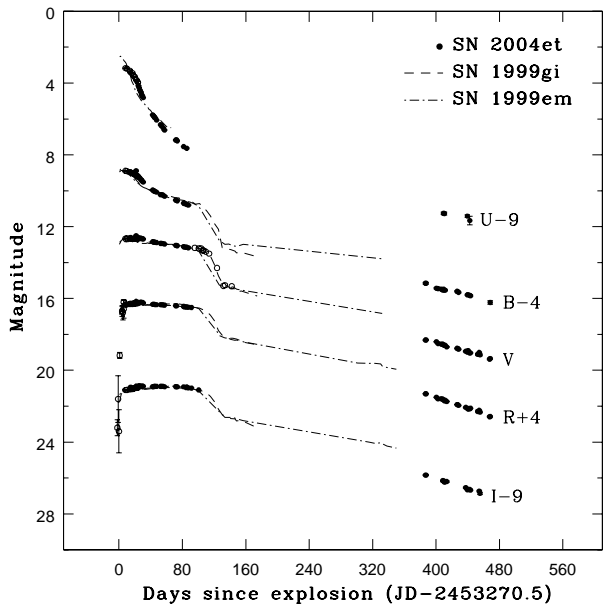}
\caption{$UBVR_{c}I_{c}$ light curve of SN 2004et (filled circle: our data), including those
available in the literature (open circle): the early data points by Li et al. (2005), the prediscovery
magnitudes by Klotz et al. (2004) and V band magnitude from Lindberg (2004).
The light curves have been shifted by arbitrary amount as indicated in the figure.
Also the light curves of SN 1999em ($UBVRI$: dash-dotted line) and SN 1999gi ($BVRI$: dash line)
have been included for comparison and shifted arbitrarily in order to match the corresponding light
curves of SN 2004et.}
\label{light_curve}
\end{figure}

The $UBVR_{c}I_{c}$ photometric data obtained from $\sim$ 14 days to $\sim$ 470 days since explosion
at 53 epochs is compiled in Table 2.
We do not have observations near the explosion epoch which are very important to study the overall
temporal evolution of the light curve. For this purpose, we have combined the early time
data points reported by Li et al. (2005) and the prediscovery magnitudes by Klotz et al. (2004)
and collaborators Yamaoka et al. (2004). A Few V band data points by Lindberg (2004)
are included at the transition from plateau to tail. The data set by Li et al. (2005) matches best with
ours as the comparison stars for the two data sets are similar, as mentioned in Section 2.
This allows us to cross compare our photometry with other data available in the literature.
U band observations, though small in number, were obtained whenever possible since they are valuable
to construct the bolometric light curve especially at early times.
The combined $UBVR_{c}I_{c}$ light curve is shown in figure \ref{light_curve}, and is typical of
type II plateau supernovae. Figure \ref{light_curve} also shows the light curves of SN 1999em
and SN 1999gi shifted arbitrarily to match those of the SN 2004et light curves.
There is a rapid decline in the U band which is followed by a slow decline
of the B band during the first 100 days after the explosion.
The decline in the B band during the first 100 days as estimated by Sahu et al. (2006) is
$\beta^{B}_{100}$ = 2.2 mag.
Patat et al. (1994) classified type II supernovae as plateau and linear on the basis of the
B band decline rates. Type IIP supernovae have $\beta^{B}_{100}$ $<$ 3.5 compared to type IIL
which have $\beta^{B}_{100}$ $>$ 3.5.
The obtained value of $\beta^{B}_{100}$ for SN 2004et establishes it as
a type IIP supernova atleast according to Patat's classification scheme.

\begin{figure}
\includegraphics[width=94mm]{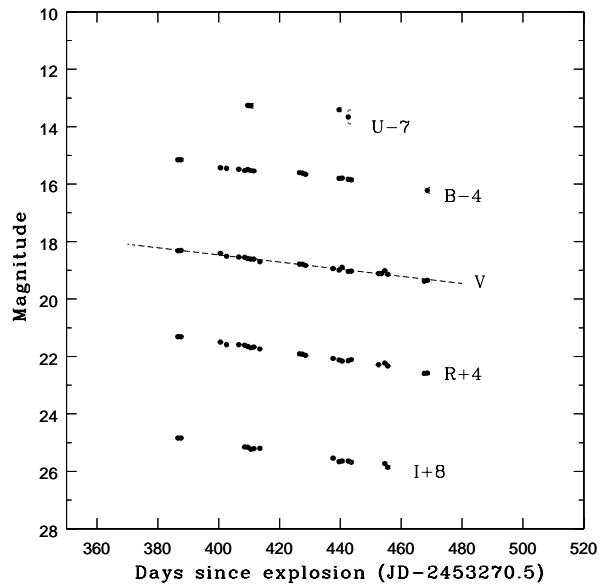}
\caption{$UBVR_{c}I_{c}$ light curve of SN 2004et during the late tail phase. The dash line indicates
the slope of 1.24 mag/(100 days) in the V band which is steeper than the expected decay of 0.98 mag/(100 days) for $^{56}$Co to $^{56}$Fe decay. The light curves have
been shifted by arbitrary amount as indicated in the figure. }
\label{tail_phase}
\end{figure}

The light curve clearly shows a pronounced plateau of constant luminosity in the $V, R_{c}, I_{c}$
bands. The plateau length is estimated to be about 110 $\pm$ 10 days from the V band light curve.
Type IIP supernovae after the
explosion, start cooling slowly to the recombination temperature of hydrogen and radiate energy
deposited into hydrogen from the initial shock. This recombination gives rise to a pronounced
plateau in the light curve during which the expansion and cooling of the photosphere balance each
other so that the luminosity remains almost constant. Immediately after the plateau phase, we
notice a steep decline in the V band light where the magnitude falls from 13.50 at 113 d to 15.32
at 142 d after the explosion. This fall of $\sim$ 2 mag in $\sim$ 30 d is clearly seen in the
V band light curve.
There is a drop in the observed flux in all other
passbands after the plateau phase. This fall indicates that the hydrogen recombination wave
has receded completely
through the massive hydrogen envelope and the supernova now enters the nebular phase.
We lack the early nebular phase observations. The decline rates stated by Sahu et al. (2006)
from $\sim$ 180 d to $\sim$ 310 d
after the explosion in the early nebular phase in B, V, R and I bands are 0.64, 1.04, 1.01 and 1.07
mag respectively.
The flux variation in this phase is marked by the radioactive decay of $^{56}$Co to $^{56}$Fe
with an expected decay rate of 0.98 mag/100 days particularly in the V band (Patat et al. 1994).
Except for the B band, the decline rates during the early nebular phase in V, R and I bands
agree fairly well with the expected decay rate of $^{56}$Co to $^{56}$Fe.
suggesting that little or no $\gamma$-rays
escaped during this time.
Figure \ref{tail_phase} shows the later nebular phase of the light curve from 386 d to 469 d
since explosion with the slope of 1.24 mag/(100 days) in the V band.
Post 370 d the decline
rate during
the later nebular phase in $BVR_{c}I_{c}$ bands is 1.01, 1.24, 1.47 and 1.49 mag/(100 days). The
decay rate beyond $\sim$ 370 d are quite steep as compared to those in the early nebular phase.
The decay rate in the nebular phase deviates
from the $^{56}$Co to $^{56}$Fe decay implying the leakage of $\gamma$-rays and the
supernova becoming transparent to the $\gamma$ rays or dust formation
in the supernova ejecta. Sahu et al. (2006) suggest dust formation in the supernova ejecta
around $\sim$ 320 d.

\subsection{Color Evolution}
\begin{figure*}
\includegraphics[width=154mm,height=154mm]{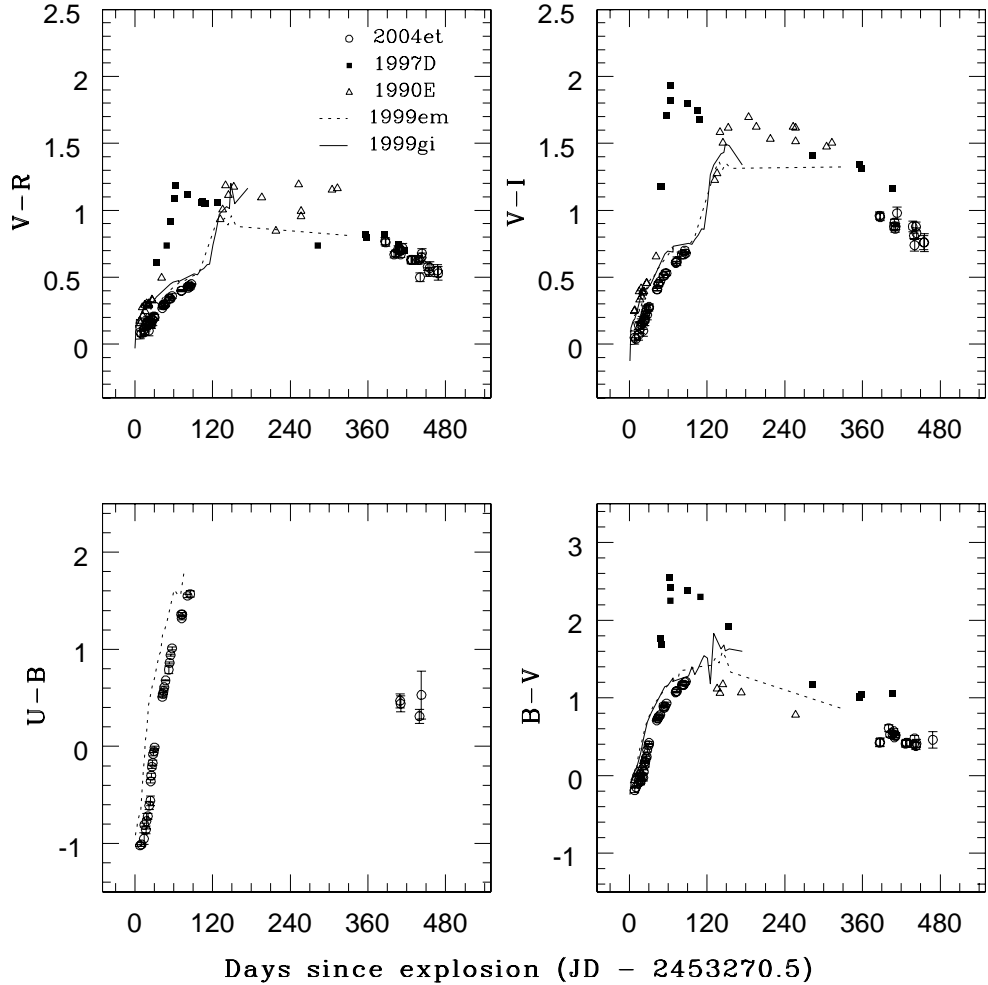}
\caption{Color evolution of SN 2004et (unfilled circle) compared with that of SN 1999em (dotted line),
SN 1999gi (solid line), SN 1997D (filled squares) and SN 1990E (unfilled triangles)
corrected for a reddening of E(B-V) = 0.41, 0.06, 0.21, 0.02 and 0.48 respectively.}
\label{color_curve}
\end{figure*}

In figure \ref{color_curve} we show the color evolution of SN 2004et along with that of SN 1999em,
SN 1999gi, SN 1997D and SN 1990E for comparison. The explosion epochs for these supernovae are
well established and taken to be JD 2451475.6 (Leonard et al. 2002a), JD 2451526.2 (Leonard et al. 2002b),
2450430.0 (Benetti et al. 2001) and 2447932.0 (Schmidt et al. 1993) for
SN 1999em, SN 1999gi, SN 1997D and SN 1990E, respectively.
The colors of these supernovae were corrected for a total reddening
(galactic + host) adopting the following E(B-V) values of 0.41, 0.06, 0.21, 0.02 and 0.48 for
SN 2004et, SN 1999em, SN 1999gi, SN 1997D and SN 1990E, respectively. The U-B and B-V color
evolution of SN 2004et is slower than that of SN 1999em which has also been studied by Li et al. (2005)
during the first month of evolution. For comparison we show the color
evolution of the proto typical faint SN 1997D. At the end of the plateau phase the colors of
SN 1997D show a sharp rise and an excess in color as compared to other supernovae. This has been
recognised to be a characteristic of faint low-luminosity supernovae such as SN 1997D, SN 1999eu
(Pastorello et al. 2004). Color evolution of SN 1990E is also quite rapid but it does not
reach a significant excess except in V-I color. Till $\sim$ 100 days after the explosion the
overall colors of SN 2004et are bluer as compared to other supernovae.
Lack of observations at the transition from plateau
to tail phase restricts us to comment on the color evolution.
Sahu et al. (2006) mention that the (B-V) and (V-R) colors of SN 2004et are similar to other SN IIP
whereas the (R-I) color of SN 2004et is still bluer as compared to SN 1999em beyond $\sim$ 200 days.
No unusual change in color is noticed around the dust formation time.

\subsection{Temporal Evolution of Photospheric Radius and Color Temperature during the Plateau Phase}
The temporal evolution of radius and color temperature is studied during the plateau phase of
SN 2004et by the Expanding Photosphere Method (EPM) summarized by Hamuy et al. (2001). The
methodology in brief is discussed in this section.
Hamuy et al. (2001) assume that if the continuum radiation arises from a spherically symmetric
photosphere, the photometric color and magnitude determines the angular radius of photosphere
given by

\begin{equation}
\theta = \frac{R}{D} = \sqrt \frac{f_\lambda}{\zeta^2_\lambda B_\lambda(T) 10^{-0.4A(\lambda)}}
\end{equation}

\noindent
where R is the photospheric radius, D is the distance to the supernova, B$_\lambda$(T) is the
Planck function at the color temperature of the blackbody radiation, f$_\lambda$ is the apparent
flux density and A($\lambda$) is the extinction. Here $\zeta_\lambda$ accounts for the fact that a
real supernova does not radiate like a blackbody at a unique color temperature. The implementation
of this requires one to make blackbody fits to the observed magnitudes. This involves the
determination of synthetic broadband magnitudes from Planck spectra. Hamuy et al. (2001),
now compute b$_{\bar\lambda}$(T), the magnitude of $\pi$B$_\lambda$(T) for a filter with central
wavelength $\bar\lambda$, to which the following polynomial is fit in the temperature range
of 4000 to 25,000 K

\begin{equation}
b_{\bar\lambda}(T) = \sum_{i=1}^5 C_{i}(\lambda) (\frac  {10^{4} K}{T})^i
\end{equation}

\noindent
The values of the coefficients C$_{i}(\lambda)$ are listed in Table 13 of Hamuy et al. (2001).
The color temperature are now computed at each epoch using these fits for any combination of
magnitude by a $\chi^2$ minimization technique. The dilution factor ($\zeta_S$) is calculated
for our photometric system by performing polynomial fits to $\zeta_{S}(T_{S})$ given by
Hamuy et al. (2001) mentioned below

\begin{equation}
\zeta_{S}(T_{S}) = \sum_{i=0}^2 a_{S,i} (\frac {10^{4} K}{T_{S}})^i
\end{equation}

\noindent
where S is the filter combination used to fit the atmosphere models with blackbody curves. The
coefficients $a_{S,i}$ for different filter combinations are listed in Table 14 of Hamuy et al. (2001).
Once the color temperature and the values of $\zeta_\lambda$ for different filter combinations
are calculated, the photospheric radius at each epoch can be obtained using equation 1 for a
known supernova distance D.

\begin{figure}
\includegraphics[width=90mm]{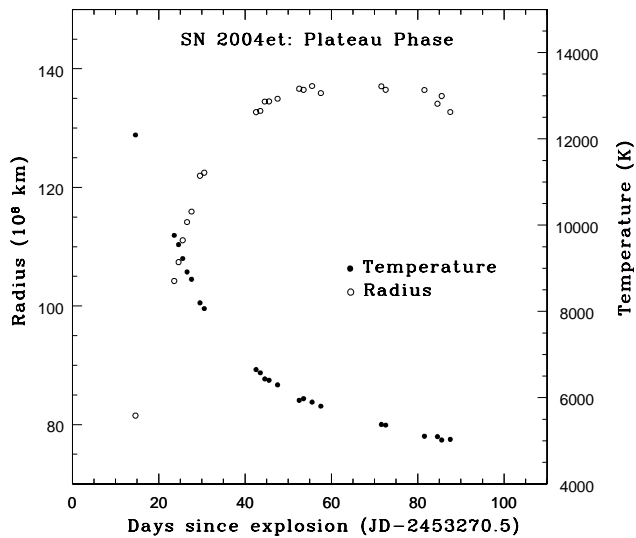}
\caption{Temporal evolution of the photospheric radius at the thermalization depth and the color
temperature for SN 2004et during the plateau phase}
\label{temp_radius}
\end{figure}

Adopting the above approach by Hamuy et al. (2001) we compute the color temperature using $BVRI$ filters and the
corresponding photospheric radius for the plateau phase of SN 2004et using
$A_{V}^{tot}$ (galactic+host) = 1.27.
The temporal evolution of color temperature and radius is shown in figure \ref{temp_radius}.
We find that the initial color temperature was $\sim$ 12100 K and the supernova cooled to $\sim$ 5000 K
till $\sim$ 90 days after the explosion.
While the photosphere was cooling during the first 90 days in the plateau phase, the
photospheric radius was increasing due to the constant expansion of the ejecta. It is
this balance between the steady increase in radius and the decrease in temperature which
very well explains the constant luminosity plateau seen in SN IIP light curves.

\subsection{The Bolometric Light Curve}
\begin{figure}
\includegraphics[width=84mm]{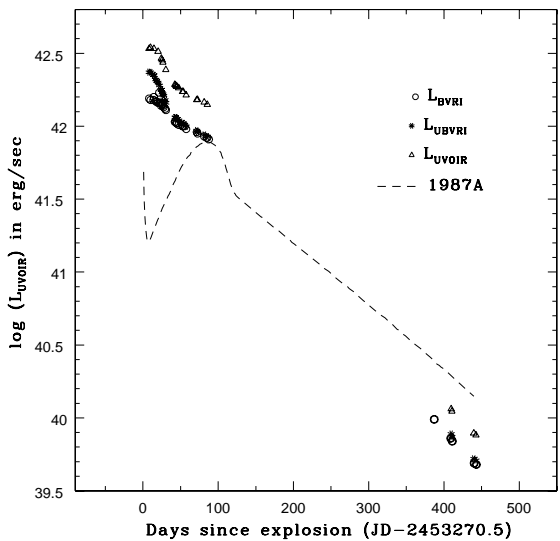}
\caption{Bolometric light curve of SN 2004et after correcting for the contribution from near IR bands
(open triangles).
For comparison we show the bolometric light curve
constructed using $UBVR_cI_c$ bands (asterisks) and that from $BVR_cI_c$ bands (open circles)
to look for the contribution of U-band flux. The dash line indicates the bolometric light curve
of SN 1987A (Suntzeff \& Bouchet 1990).}
\label{bolometric_sn2004et_ubvri_bvri}
\end{figure}

\begin{figure}
\includegraphics[width=114mm]{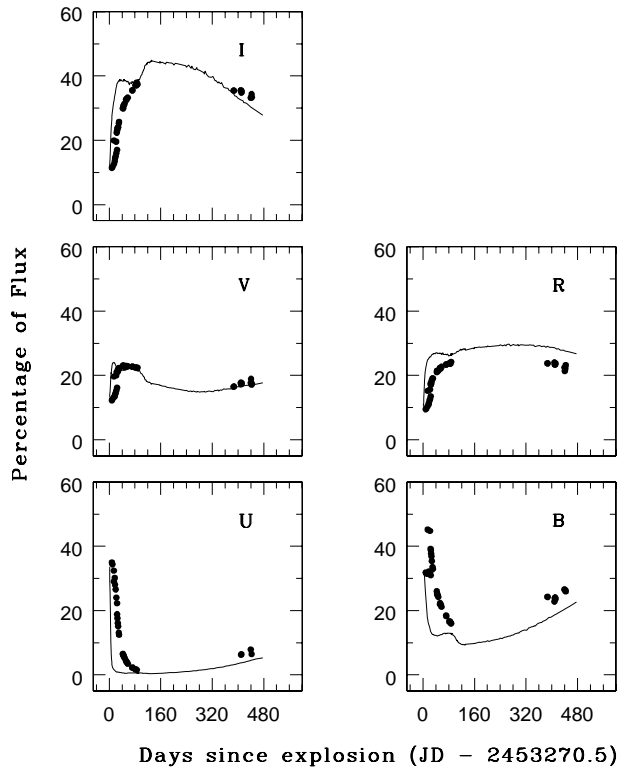}
\caption{Flux contribution, in percentage, in UBVR$_c$I$_c$ bands of SN 2004et along with a
comparison to SN 1987A}
\label{contrib_flux}
\end{figure}

The integrated flux in $UVOI_cR_c$ bands gives a meaningful estimate of the bolometric luminosity.
The bolometric luminosity in the tail phase provides an accurate estimate of the $^{56}$Ni mass
ejected in the explosion. The distance to the host galaxy NGC 6946 is well estimated using different
methods such as the HI Tully-Fisher relation (Pierce 1994), the CO Tully-Fisher relation
(Schoniger \& Sofue 1994) and the Expanding Photosphere method (EPM) for type II SNe
(Schmidt et al. 1994).
We adopt a mean distance of 5.5 $\pm$ 1.0 Mpc. To construct the bolometric light curve, we consider
here the epochs which have observations in all $UBVR_cI_c$ bands. Our data set is supplemented by
the early time $UBVR_cI_c$ data by Li et al. (2005). The UVOIR bolometric light curve is thus
constructed using the de-reddened magnitudes and the known distance to NGC 6946. The correction for the
missing IR flux has been applied by comparing with the bolometric light curve for SN 1987A. The redshift of
the host galaxy NGC 6946 is very small (z = 0.00016; from NED), we have therefore neglected the
k-correction. The obtained magnitudes were converted to flux using calibrations by Bessel et al. (1998).
The U-band contribution at $\sim$ 8 day is $\sim$ 35 percent and that from the I$_c$-band is
$\sim$ 11 percent. The contribution from the U band decreases towards the end of the plateau phase
whereas that from the I band increases. The similar trend is there for SN 1987A. In figure
\ref{contrib_flux} we show the percentage contribution of flux in different bands of SN 2004et
during the plateau and the late nebular phases. For a comparison, percentage contribution of
flux in the case of SN 1987A in different bands has also been shown. We notice that
during the late nebular phase the contribution from the U band starts to increase and is
$\sim$ 5 percent and that from I$_c$ band decreases and reached $\sim$ 28 percent at $\sim$ 476 d.
The U band contribution during the initial days is significant as compared to I$_c$ and R$_c$
bands. We have taken into account any contribution from the near IR bands while
constructing the UVOIR bolometric light curve.
In figure \ref{bolometric_sn2004et_ubvri_bvri} we show the
UVOIR bolometric light curve constructed from $UBVR_cI_c$ bands as asterisks. The open circles
in the figure shows the contribution from the $BVR_cI_c$ bands alone, showing that the U-band
flux is dominant during the early phase. The open triangles show the $UVOI_cR_c$ bolometric light curve. The
figure clearly shows that we miss out quite a lot of flux if we do not account for the contribution from
the near IR bands.
The bolometric luminosity
of the exponential tail gives an estimate of $^{56}$Ni mass ejected during the explosion by a
direct comparison to the bolometric light curve of SN 1987A (shown by dashed line in figure
\ref{bolometric_sn2004et_ubvri_bvri}). The estimation of $^{56}$Ni mass using the bolometric light
curve is discussed in detail in the next section.

\subsection{Physical Parameters}
\subsubsection{Estimate of ejected Nickel mass}
The tail phase of the light curve of a type IIP supernova is thought to be powered by the
radioactive decay of $^{56}$Co to $^{56}$Fe. The luminosity in the tail phase is directly
proportional to the amount of nickel produced in the explosion. To estimate the amount of
nickel produced in the explosion of SN 2004et, we use here different methods.
These methods are briefly described below.

\begin{itemize}
\item {\it Nickel mass from the bolometric luminosity of the exponential tail}\\
The nickel mass derived from the bolometric luminosity of the exponential tail as described
by Hamuy (2003) assumes that all the $\gamma$-rays during the radioactive decay of $^{56}$Co
to $^{56}$Fe are fully thermalized. It is to be noted here that $^{56}$Co is the daughter
nucleus of $^{56}$Ni, thus the bolometric luminosity in the exponential tail is proportional
to the ejected $^{56}$Ni mass. The late time decline rates for SN IIP are consistent with the
decay of $^{56}$Co to $^{56}$Fe.
The V band ($V_{t}$) magnitude in the exponential tail is
first converted into bolometric luminosity ($L_{t}$) in erg/sec using a total extinction of
$A_{V}$ = 1.27, a distance of 5.5 $\pm$ 1.0 Mpc and applying a bolometric correction of
BC = 0.26 $\pm$ 0.06. The nickel mass is then estimated by using eq. (2) of Hamuy (2003).
We estimate the nickel mass using this method at two different points in the tail and obtain an
average value of nickel ejected in the explosion as $M_{Ni}$ = 0.06 $\pm$ 0.03 $M_{\odot}$.

\item {\it Nickel mass from `steepness of decline' S correlation}\\
Elmhamdi et al. (2003) present a correlation between the rate of decline from plateau to tail
in the V band and the nickel mass.
Elmhamdi et al. (2003) define a steepness `$S$' parameter, which is the maximum gradient
during transition in mag/day. Elmhamdi et al. (2003) using a sample of type IIP supernova conclude
that the steepness `$S$' anticorrelates with the $^{56}$Ni mass. The smaller the amount of $^{56}$Ni,
the steeper will be the transition from the plateau to tail.
For SN 2004et, we do not have a well sampled V band light curve during the transition from the
plateau to the tail. Hence the determination of $S$ will result in large uncertainties.
Using the available data set we obtain S = 0.07 $\pm$ 0.02 which results in
$M_{Ni}$ = 0.056 $\pm$ 0.016 M$_\odot$ using eq. (3) of Elmhamdi et al. (2003).

\end{itemize}

Sahu et al. (2006) also estimate the nickel mass by comparing the tail phase bolometric luminosity
with that of SN 1987A during $\sim$ 250 d to 300 d if we assume that the $\gamma$ ray deposition for both
SN 2004et and SN 1987A is the same. The estimated mass of $^{56}Ni$ is 0.048 $\pm$ 0.01 M$_\odot$
for a value of 0.075 M$_\odot$ for SN 1987A (Turatto et al. 1998).
But we lack the early nebular phase observations and towards the late nebular phase the $\gamma$
ray deposition for SN 2004et and SN 1987A is not really the same and there is evidence of dust
formation around 310 days. Thus, we do not adopt this method for nickel mass estimation.

We discussed here different methods to determine the amount of nickel ejected in the explosion and see
that values obtained are consistent with each other within errors.

\subsubsection{Progenitor star properties}
Litvinova \& Nad\"{e}zhin (1985) investigated the course of type II supernovae by constructing
hydrodynamical models. They investigated the dependence of supernova outburst on three basic
parameters: the ejected mass of the envelope (M$_{ej}$), the pre-supernova radius (R$_{0}$)
and the energy of explosion (E). Litvinova \& Nad\"{e}zhin (1985) obtained approximate
expressions for M$_{ej}$, R$_{0}$ and E in terms of three observational parameters: the
plateau duration (t$_{p}$), the absolute V magnitude at the mid plateau epoch (M$_{V}$) and
the expansion velocity of the photosphere at the mid plateau epoch (V$_{ph}$). Similarly,
Popov (1993) derived somewhat different expressions for M$_{ej}$, R$_{0}$ and E based on the
analytical models. The input parameters required in the model are the plateau duration (t$_{p}$),
the photospheric velocity (V$_{ph}$) and the absolute magnitude (M$_{V}$).

The detailed photometric observations of SN 2004et show that the
plateau length (t$_{p}$) was 110 $\pm$ 10 days and the absolute V magnitude (M$_{V}$) at
the mid of the plateau is -17.08 $\pm$ 0.39.
The spectroscopic observations of weak iron lines indicate a mid plateau velocity (V$_{ph}$)
of 3560 $\pm$ 100 km/sec as reported by Sahu et al. (2006).
We use the photospheric velocity given by Sahu et al. (2006) at the mid plateau epoch for further analysis.
Using the relations by Litvinova \& Nad\"{e}zhin (1985) and Popov (1993) we derive
the parameters M$_{ej}$, R$_{0}$ and E for SN 2004et which are listed in
Table \ref{progenitor_properties}.
The obtained estimates of M$_{ej}$, R$_{0}$ and E using expressions by
Litvinova \& Nad\"{e}zhin (1985) and Popov (1993) are consistent within errors.

\begin{table}
\caption{Progenitor star parameters}
\smallskip
\begin{center}
\begin{tabular}{@{}llll}\hline \hline
&Ejected & Pre-supervova & Explosion\\
& Mass & Radius & Energy\\
&M$_{ej}$ (M$_{\odot}$) & R$_{0}$ (R$_{\odot}$) & E ($\times 10^{51}$) (erg)\\
\hline
&&& \\
Litvinova & 16 $\pm$ 5&530 $\pm$ 280 &0.98 $\pm$ 0.25\\
$\&$ Nad\"{e}zhin & & &\\
Popov & 8 $\pm$ 4&1251 $\pm$ 937 &0.60 $\pm$ 0.32 \\
&&& \\ \hline
\end{tabular}
\end{center}
\label{progenitor_properties}
\end{table}

If approximately 1.5 M$_{\odot}$ is enclosed in the neutron star and 0.5 - 1.0 M$_{\odot}$ is
lost by the wind, we obtain a range of roughly 10 to 20 M$_{\odot}$ for the progenitor of SN 2004et.
Sahu et al. (2006) find that the [OI] luminosities of SN 2004et and SN 1987A are comparable. The [OI]
luminosity of SN 1987A corresponds to an Oxygen mass in the range 1.5 - 2 M$_{\odot}$. This Oxygen mass
corresponds to a main sequence mass of 20 $M_{\odot}$ for SN 2004et.
Li et al. (2005) derive the progenitor mass by analysing the high-resolution CFHT pre-supernova images of
NGC 6946 and suggest the progenitor to be a yellow supergiant of 15$^{+5}_{-2}$ M$\odot$.
Further details on the evolutionary sequence of the progenitor can be obtained from Li et al. (2005).
The progenitors of SN IIP are generally considered to be red supergiants with thick hydrogen
envelopes which gives rise to the plateau phase in the optical light curve and a typical P-Cygni
spectral profile. The spectroscopic analysis at 9 and 20 days after the explosion carried out by
Li et al. (2005) shows a lack of a typical P-Cygni profile.
Another interesting feature of SN 2004et to note here is the radio emission detected
on October 05, 2004 (Stockdale et al. 2004) just 14 days after the explosion which clearly suggests the
presence of dense circumstellar material. Therefore, it is likely that the progenitor experienced a
RSG phase close to the time of explosion. Chevalier et al. (2006) modelled the radio data and found
a progenitor mass of $\sim$ 20 M$\odot$ for SN 2004et which includes the mass loss rate obtained
from radio observations.
The measurements of the progenitor star by Li et al. (2005) were teken at ground based resolution and
it is possible that the star observed was blended with another star that did not partake in the explosion.
The progenitor masses for SN 2004et obtained using different methods is consistent with the results
within uncertainties. The light curves and the pre-supernova radius \ref{parameter_SNIIP_sample}
derived here suggests that the progenitor of SN 2004et is likely a red supergiant.

\begin{table*}
\caption{Parameters of Supernovae IIP sample (*)}
\smallskip
\begin{center}
\begin{tabular}{@{}lllllllllll}\hline \hline
&Parent & Distance & $A_V^{tot}$ & M$^{V}$ & $t_{p}$ & E (10$^{51})$ & R (R$_{\odot}$) & M$_{ejected}$ & M$_{ms}$ & Estimated\\
Supernova & Galaxy & & & & & & & & & $^{56}$Ni\\
& & (Mpc) & (mag) & (mag) & (days) & (ergs) & & (M$_{\odot}$) & (M$_{\odot}$) &(M$_{\odot}$)\\
\hline
&&&&&&&&&&\\
1987A & LMC & 0.05 & 0.60 & & 40 & 1.3 & 40 & 15 & 20 & 0.075\\
&&&&&&&&&&\\
1990E & NGC 1035 & 21.0$\pm$3 & 1.5 & -16.93 & 131$\pm$10 & 3.4$^{+1.3}_{-1.0}$ & 162$^{+148}_{-78}$ & 48$^{+22}_{-15}$ & & 0.073$^{+0.018}_{-0.051}$\\
&&&&&&&&&&\\
1997D & NGC 1536 & 13.43 & 0.07 & -14.65 & 50 & 0.1 & 85 & 6$\pm$1 & 8-12 & 0.002\\
&&&&&&&&&&\\
1999em & NGC 1637 & 8.2$\pm$0.6 & 0.31 &-16.48 & 95 & 1.2$^{+0.6}_{-0.3}$ & 249$^{+243}_{-150}$ & 27$^{+14}_{-8}$ & 12$\pm$1 & 0.042$^{+0.027}_{-0.019}$\\
&&&&&&&&&&\\
1999gi & NGC 3184 & 11.1 & 0.65 & -15.68 & 95 & 1.5$^{+0.7}_{-0.5}$ & 81$^{+110}_{-51}$ & 43$^{+24}_{-14}$& 15$^{+5}_{-3}$ & 0.018$^{+0.013}_{-0.009}$\\
&&&&&&&&&&\\
2003gd & M74 & 9.3$\pm$1.8 & 0.43 & -15.92 & 67$^{+34}_{-25}$ & & & 6 & 8$^{+4}_{-2}$ & 0.016 $^{+0.010}_{-0.006}$\\
&&&&&&&&&&\\
2004A & NGC 6207 & 20.3$\pm$3.4 & 0.19 & -16.24 & 80$^{+25}_{-5}$ & & & 11$^{+10}_{-4}$ & 9$^{+3}_{-2}$ & 0.046 $^{+0.031}_{-0.017}$\\
&&&&&&&&&&\\
2004dj & NGC 2403 & 3.47$\pm$0.2 & 0.22 & -15.88 & 100$\pm$20 & 0.86$^{+0.89}_{-0.49}$ & 155$^{+150}_{-75}$ & 19$^{+20}_{-10}$ & $>$20 & 0.02$\pm$0.01\\
&&&&&&&&&&\\
{\bf 2004et} & {\bf NGC 6946} & {\bf 5.5$\pm$1.0} & {\bf 1.27} & {\bf -17.08} & {\bf 110$\pm$10} & {\bf 0.98$\pm$0.25}&{\bf 530$\pm$280} & {\bf 16$\pm$5} & {\bf $\sim$ 20} &{\bf 0.06$\pm$0.03} \\
&&&&&&&&&&\\
2005cs & M51 & 8.4 & 0.34 & -15.2 & 90-120 & & & & 7-12 & $\leq$ 10$^{-2}$\\
&&&&&&&&&&\\
\hline
\end{tabular}
\end{center}
\label{parameter_SNIIP_sample}
* References: SN 1987A: Hamuy et al. 2003, 
Woosley et al. 1989;
SN 1990E: Schmidt et al. 1993, Hamuy et al. 2003;
SN 1997D: Toratto et al. 1998, Chugai et al. 2000;
SN 1999em: Leonard et al. 2002, Hamuy et al. 2003;
SN 1999gi: Leonard et al. 2002, Hamuy et al. 2003;
SN 2003gd: Hendry et al. 2005;
SN 2004A: Hendry et al. 2006;
SN 2004dj: Vinko et al. 2006;
SN 2004et: Present Work (the progenitor star properties are those obtained
from Litvinova \& Nad\"{e}zhin (1985));
SN 2005cs: Pastorello et al. 2006.
\end{table*}

\section{A comparison with other SN IIP}
\begin{figure*}
\includegraphics[width=154mm,height=154mm]{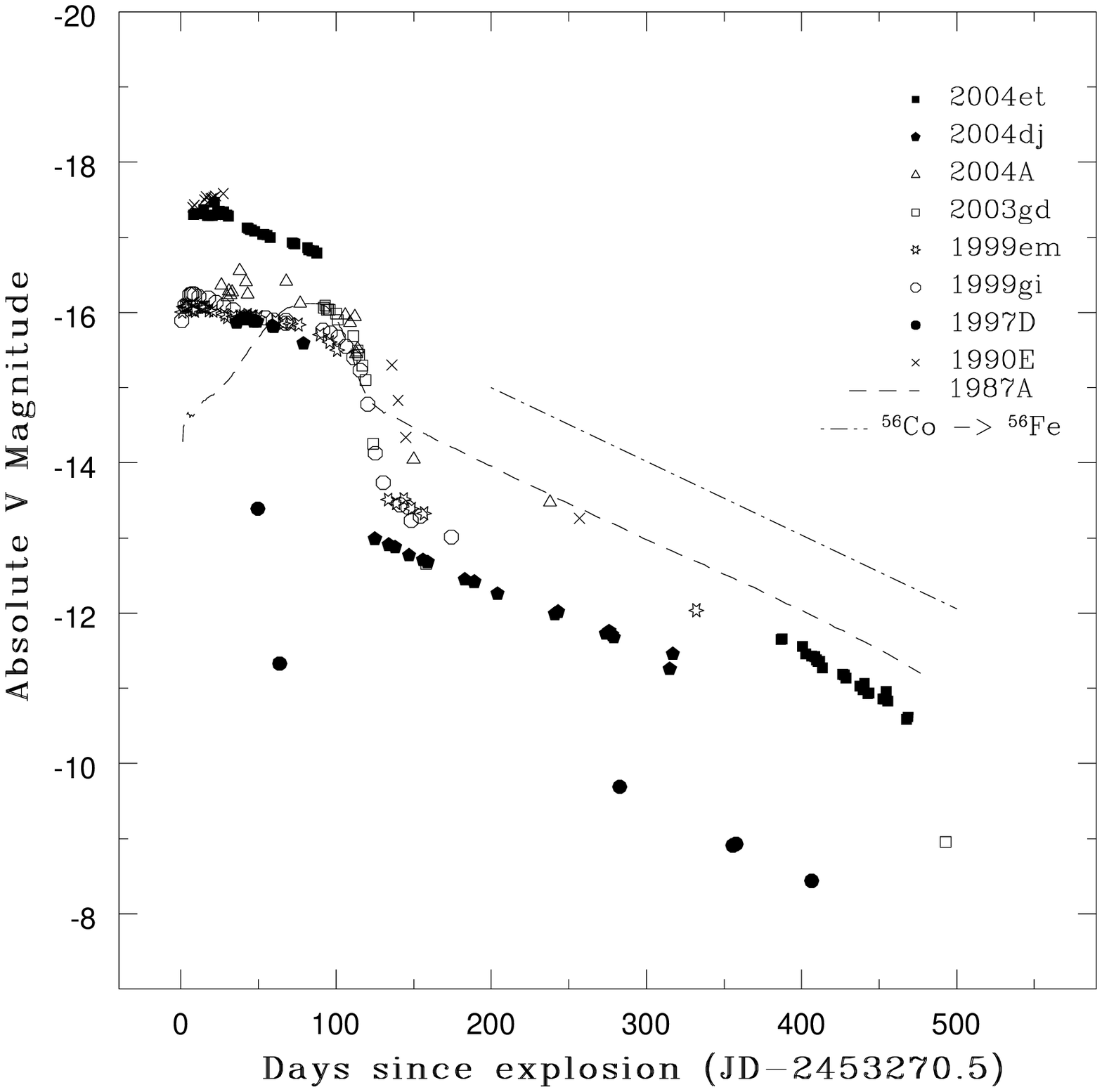}
\caption{Comparison of absolute V band light curves of SN 2004et, SN 2004dj, SN 2004A, SN 2003gd,
SN 1999gi, SN 1999em, SN 1997D, SN 1990E and SN 1987A. The corresponding symbols representing
each supernova are shown in the figure. The magnitudes have been corrected for reddening and
distance listed in Table \ref{parameter_SNIIP_sample}.}
\label{abs_Vmag}
\end{figure*}

The light curves of SN 2004et are similar to those of typical type IIP supernovae SN 1999em and
SN 1999gi during the plateau phase shown in figure \ref{light_curve}. During the transition,
the V band light curve of SN 2004et matches well with SN 1999gi. Color evolution
of SN 2004et is studied in figure \ref{color_curve}
and we see that the U-B and B-V colors of SN 2004et evolves slowly than that of SN 1999em.
SN 2004et appears to be more like a normal type IIP apart from the difference in U-B and
B-V color evolution with SN 1999em.
The spectrum of SN 2004et
has a bluer continuum than SN 1999em (Li et al. 2005). The peculiar P-Cygni profile in H$\alpha$ is dominated by the
emission component unlike the typical P-Cygni profile in SN 1999em. SN 2004et evolves slower than
SN 1999em in the UV part of the spectrum which explains the slow photometric evolution of U-B and
B-V colors. H\"{o}flich et al. (2001) based on theoretical models find that though the peak
luminosity varies among SN IIP, but the average absolute brightness during the plateau phase
remains more or less the same with $\bar{M}_{V}$ $\approx$ -17.6 $\pm$ 0.6 mag.
In figure \ref{abs_Vmag}
we compare the absolute V band light curves
of SN 2004et with those of SN 2004dj, SN 2004A,
SN 2003gd, SN 1999em, SN 1999gi, SN 1997D, SN 1990E and SN 1987A. The parameters of these supernovae
along with references are mentioned in Table \ref{parameter_SNIIP_sample}.
The absolute V band light curve of SN 2004et during
the plateau phase matches very well with that of SN 1990E. The bolometric luminosities for these two
supernovae are similar during the plateau phase but the tail luminosity of SN 2004et is lower
than that of SN 1990E. This is because of the lower $^{56}$Ni mass synthesised during the explosion of
SN 2004et as compared to SN 1990E.

Chevalier et al. (2006) place the supernovae in two mass groups, 8 - 13 M$\odot$
and 13 - 18 M$\odot$, the low mass and the high mass group respectively. The progenitor mass of
different type IIP supernovae are listed in Table \ref{parameter_SNIIP_sample}. There exist clearly
two groups of supernovae with high mass progenitors (SN 1999em, SN 1999gi, SN 2004dj, SN 2004et)
and low mass progenitors (SN 1997D, SN 2003gd, SN 2004A, SN 2005cs). Thus, placing SN 2004et progenitor
in the upper end of the mass range of SN IIP.

\section{X-ray emission}

We interpret the X-ray emission in terms of the circumstellar interaction model for type IIP SNe
discussed by Chevalier et al.\ (2006) and references therein.  The first two observations
have temperatures above $\sim$\ee{5}{7}~ K (and the last is only slightly below this), a regime in which
the reverse shock is expected to be nonradiative and free-free emission dominates.  The X-ray luminosity
is given by their eq.~(9), which comes from Fransson, Lundqvist, \& Chevalier (1996):
\begin{eqnarray*}
&\frac{dL_\mathrm{rev}}{dE} = 2\times10^{35}\zeta(n-3)(n-4)^2 T_8^{-0.24}e^{-0.116/T_8}&\\
&~~~~~~~~~~~~\times\left(\frac{\dot{M}_{-6}}{v_{w1}}\right)^2 V_{s4}^{-1}\left(\frac{t}{10~\mathrm{days}}\right)^{-1}~\mathrm{ergs\ s}^{-1}~\mathrm{keV}^{-1}.&
\end{eqnarray*}
where $\zeta$ is 0.86 for solar abundances, $n$ is the index of the ejecta density profile
($\rho_\mathrm{SN} = A t^{-3} (r/t)^{-n}$), $T_8$ is the temperature in units of $10^8$ K,
$\dot{M}_{-6}$ is the mass loss rate of the progenitor in units of
 $10^{-6}~M_\odot~\mathrm{yr}^{-1}$, $v_{w1}$ is the wind velocity in units of 10~\kms, $V_{s4}$
is the shock velocity in units of $10^4~\kms$, and $t$ is the time since explosion.

Following the discussion of 2004et in Chevalier et al. (2006), we take $n=10$ and use an
expansion velocity of 15,000~\kms\ at 10 days (based on the maximum velocity of
14,200~\kms\ found by Li et al.\ 2005).  The shock velocity is expected to evolve as
$t^{-0.10}$ (Chevalier et al.\ 2006).  The typical wind velocity for a yellow supergiant
is not well known; red supergiants are expected to have winds with velocities of 10--15~\kms.

We use each of our three X-ray measurements to estimate the mass-loss of the
progenitor, and find values of $(\dot{M}_{-6}/v_{w1})$ = 2.2, 2.1, and 2.5,
respectively, assuming that the reverse shock is responsible for the X-ray
emission.  We can check the expected strength of the circumsteller shock
using eq. (8) of Chevalier et al.\ (2006).  This predicts \Lx\ (over the
0.5--8 keV band) of 2.9, 0.8, and $0.5\times10^{36}$~erg~s$^{-1}$,
indicating that the reverse shock does indeed dominate.

However, there are some discrepancies between the data and this model.  For
example, the expected evolution of the total X-ray luminosity when free-free
emission dominates as $\Lx\propto t^{-1}$ (Chevalier \& Fransson 1994), much
different than the $t^{-0.4}$ evolution seen.  One possibility is that the
harder band (2--8 keV) X-ray emission is due to the reverse shock, and it
shows roughly the expected evolution (Figure~\ref{fig:xraylc}).  The softer
band (0.5--2 keV) emission represents some (unexplained) steady component.
Under this assumption, we calculate the mass-loss using only the 2--8 keV
emission and find $(\dot{M}_{-6}/v_{w1})$ = 2.1, 1.8, and 2.0 for each of the
three observations respectively (again for $n=10$).
The temperature evolution is much more rapid than expected in the reverse shock model, in which
 $T\propto V_{s4}^2 \propto t^{-0.2}$ for $n=10$ (Fransson et al.\ 1996). However, if there are two
distinct components of the emission (a steady soft component and a decaying hard component), these simple
 single-temperature models may be misleading.  In addition, a single-temperature model of just the emitting
shock region may be inaccurate (e.g., see the discussion in Nymark, Fransson, \& Kozma 2006).

A faster temperature evolution is expected for a smaller value of $n$, but even $n=7$ results in a much slower
evolution ($T\propto t^{-0.4}$) than is seen.  However, if a smaller $n$ is part of the explanation of the rapid
temperature evolution, the mass loss estimates would increase.  For $n=7$, we calculate
 $(\dot{M}_{-6}/v_{w1})$ = 5.3, 6.0, and 7.5 for each of the three observations, respectively.

\section {Radio emission at 1.4 GHz}
SN 2004et was detected at radio wavelengths by
Stockdale et al. (2004) using the VLA on 2004 October 5.128 UT which is just
$\sim$ 14 days after the explosion. It was extensively monitored at 22.5 GHz,
14.9 GHz, 8.46 GHz (Stockdale et al. 2004) and 4.99 GHz (Beswick et al. 2004) .
Chevalier et al. (2006) present the data in other frequencies along with the modelled
light curve. We have the lowest frequency observation available here at 1.39 GHz using the
GMRT.

\begin{figure}
\includegraphics[width=94mm]{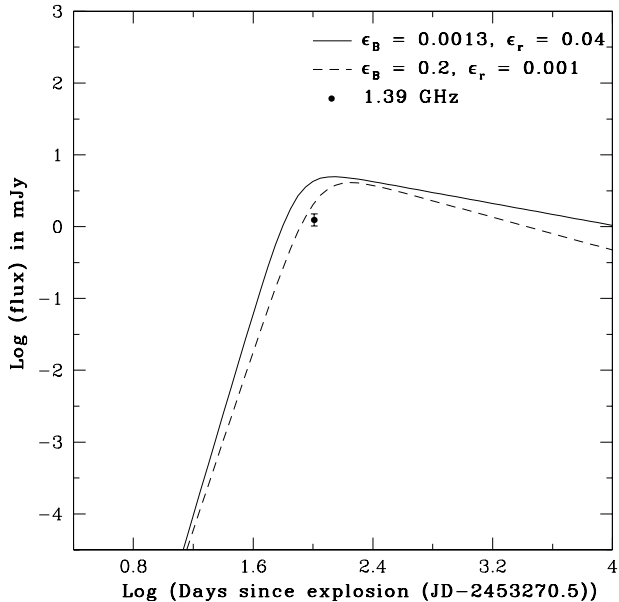}
\caption{Observed radio data point of SN 2004et at 1.39 GHz compared to the model
given by Chevalier et al. (2006) with two different combinations of $\epsilon_B$ and
$\epsilon_r$}
\label{generated_1.4}
\end{figure}

Chevalier et al. (2006) have presented modelled light curves at different radio frequencies
using two different combinations of $\epsilon_B$ and $\epsilon_r$, where $\epsilon_B$ is the
fraction of total blast wave energy in post shock magnetic field and $\epsilon_r$ that in relativistic
particles.
The two combinations are $\epsilon_B$ = 0.0013, $\epsilon_r$ = 0.04 and
$\epsilon_B$ = 0.2, $\epsilon_r$ = 0.001. The
details of the model can be obtained from Chevalier et al. (2006). We have scaled the radio
light curves at different frequencies presented by Chevalier et al. (2006) to 1.39 GHz
frequency. The scaled light curve obtained for 1.39 GHz for the two different combinations
of $\epsilon_B$ and $\epsilon_r$ are shown in figure \ref{generated_1.4}.
We see that the observed flux at 1.39 GHz is close to that expected in both light curves, but it is
in somewhat better agreement with that for the combination $\epsilon_B$ = 0.2, $\epsilon_r$ = 0.001.

Based on the 2--8 keV luminosity of X-ray observations adopting n = 10, we estimate a mass-loss rate
for the progenitor star of
$\sim$\ee{2}{-6}~$M_\odot~\mathrm{yr}^{-1}$ for an assumed wind velocity of 10~\kms similar to the
standard mass loss prescription of $M_{-6}$ = 1.3 -- 3 for a progenitor mass of 15 $M_\odot$.
The value of $(\dot{M}_{-6}/v_{w1})$ is determined by the turn on time but it also depends on the
temperature of the circumstellar medium. The radio model of Chevalier et al. (2006) lead to an
estimate of $(\dot{M}_{-6}/v_{w1})$ $\approx$ 10$T_{cs5}^{3/4}$. With $(\dot{M}_{-6}/v_{w1})$ as determined
from X-ray observations, we deduce that the temperature of the unshocked circumstellar medium (CSM) is
T$_{cs}$ $\sim$ 10$^4$ K.

\section{Conclusions}
The conclusions of the present paper are summarized below:
\begin{itemize}
\item
We present the \chandra\ X-ray observations of a type IIP supernova SN 2004et at three
epochs. We extract the X-ray spectra and try to fit two simple models -- an absorbed power law
and an absorbed mekal plasma model. We notice that the spectra softens with time. The
X-ray luminosity is well characterised by $\Lx \propto t^{-0.4}$.

\item
An extensive photometric coverage starting from $\sim$ 14 to 470 days since explosion is presented.
The explosion date is well constrained at Sept. 22.0, 2004 (JD 2453270.5) based on
the pre-discovery magnitudes which is about 5 days before the discovery on Sept. 27.0, 2004.
SN 2004et showed a pronounced plateau phase with a
plateau duration of $\sim$ 110 $\pm$ 10 days. The average absolute V magnitude during the
plateau phase is -17.08. The light curves and color curves of SN 2004et were compared with
those of SN 1999em and SN 1999gi. The U-B and B-V color evolution of SN 2004et is different
than SN 1999em and they evolve slowly compared to SN 1999em.

\item
Temporal evolution of photospheric radius and color temperature is studied during the plateau phase
using the technique by Hamuy et al. (2003). The temperature decreases from 12100 K to 5000 K with an
increase in the photospheric radius.

\item
The bolometric light curve of SN 2004et is constructed by integrating the flux in $UBVR_{c}I_{c}$ bands
using a distance estimate of 5.5 $\pm$ 1.0 Mpc to NGC 6946 and the known reddening of
E(B-V) = 0.41 $\pm$ 0.07. The tail bolometric luminosity was compared to that of SN 1987A.

\item
The ejected $^{56}$Ni mass estimated in the explosion is 0.06 $\pm$ 0.03 M$_{\odot}$.
The physical parameters such as the ejected mass (M$_{ej}$), the pre-supernova radius (R$_{0}$)
and the explosion energy (E) are estimated using the expressions given by
Litvinova \& Nad\"{e}zhin (1985) based on hydrodynamical models and those given by Popov (1993)
based on the analytical models. This gives the ejected mass in the range 8 $-$ 16 M$_{\odot}$ and
a progenitor star of $\sim$ 20 M$_{\odot}$.

\item
The radio observations at 1.4 GHz were carried out from GMRT. The 1.4 GHz observation was compared with the
predictions of the model proposed by Chevalier et al. (2006). Thus, the model by Chevalier et al. (2006)
reproduces and explains the light curve at 1.4 GHz, fairly.

\item
The bulk of the X-ray emission detected by \chandra\ is due to a reverse shock.  The 2--8 keV luminosity roughly follows the expected evolution of free-free emission from the reverse shock.  The 0.5--2 keV luminosity stays roughly constant.  Based on the 2--8 keV luminosity, we estimate a mass-loss rate for the progenitor star of $\sim$\ee{2}{-6}~$M_\odot~\mathrm{yr}^{-1}$ for an assumed wind velocity of 10~\kms.

\end{itemize}

\section*{ Acknowledgement}
We thank all the observers at Aryabhatta Research Institute of Observational Sciences (ARIES)
who granted their valuable time and support for the continous observations of this event over
a span of almost one and a half years. We are thankful to GMRT staff for carrying out our observations.
GMRT is operated by the National Center for Radio
Astrophysics of the Tata Institute of Fundamental research. We thank the anonymous referee for useful 
comments which helped to improve the manuscript. 
IRAF is distributed by the National Optical Astronomy Observatories, which are operated
by the Association of Universities for Research in Astronomy, Inc., under contract to the
National Science Foundation. AIPS is run by National Radio Astronomy Observatory (NRAO).  D.~P.\ gratefully acknowledges the support provided by NASA through Chandra Postdoctoral Fellowship grant PF4-50035 awarded by the Chandra X-Ray Center, which is operated by the Smithsonian Astrophysical Observatory for NASA under contract NAS8-03060.
WHGL acknowledges support from Chandra.

\end{document}